\documentclass{article}
\usepackage{amsmath}

\begin{document}

\begin{center}

TWO-BRANE RANDALL-SUNDRUM MODEL IN $AdS_5$ AND $dS_5$\\
\bigskip
I. Brevik\footnote{Department of Energy and Process Engineering, Norwegian University of Science and Technology, N-7491 Trondheim, Norway. E-mail: iver.h.brevik@mtf.ntnu.no}, K. B{\o}rkje\footnote{Department of Physics, Norwegian University of Science and Technology, N-7491 Trondheim, Norway. E-mail: kjetil.borkje@phys.ntnu.no}, J. P. Morten\footnote{Department of Physics, Norwegian University of Science and Technology, N-7491 Trondheim, Norway. E-mail: jan.morten@phys.ntnu.no}
\end{center}
\bigskip
\bigskip
\begin{abstract}
Two flat Randall - Sundrum three-branes are analyzed, at fixed mutual distance, in the case where each brane contains an ideal isotropic fluid. Both fluids are to begin with assumed to obey the equation of state  $p=(\gamma -1)\rho$, where  $\gamma$ is a constant. Thereafter, we impose the condition that there is zero energy flux from the branes into the bulk, and assume that the tension on either brane is zero. It then follows  that constant values of the fluid energies at the branes are obtained only if the value of $\gamma$ is equal to zero (i.e., a `vacuum' fluid). The fluids on the branes are related: if one brane is a $dS_4$ brane (the effective four-dimensional constant being positive), then the other brane is $dS_4$ also, and if the fluid energy density on one brane is positive, the energy density on the other brane  is larger in magnitude but {\it negative}. This is a non-acceptable result, which sheds some light on how far it is possible to give a physical interpretation of the two-brane scenario. Also, we discuss the graviton localization problem in the two-brane setting, generalizing prior works.
\end{abstract}
\begin{center}
February 2004
\end{center}

\newpage
\section*{\label{I} I. Introduction}

Consider a flat Randall - Sundrum (RS) three-brane \cite{randall99} situated at the position $y=0$ in the transverse $y$ direction. Assume that on the brane there is an isotropic ideal fluid, obeying the equation of state $p= (\gamma -1)\rho$, with $\gamma$ a constant. Brane dynamics of such a configuration - as well as of the analoguous two-brane configuration - has been analyzed extensively in several papers \cite{binetruy00,binetruy00a,langlois00,binetruy01,langlois02,langlois03,langlois03a,binetruy03} (also with quantum corrections \cite{nojiri01}). The purpose of the present study is to focus attention on the following two points in the description of the two-brane system:

(1)  Assuming a fixed interbrane distance $R$ we wish to analyze, after having determined the components of the five-dimensional metric, to what extent the fluids on the two branes are dependent on each other. For simplicity, we set the brane tensions $\sigma$ equal to zero. Also, we assume zero energy flux in the $y$ direction. It turns out that, in order to preserve time independence of the energy densities of the two fluids,  one has to impose the condition of a "vacuum"  fluid, $p=-\rho$,  on each brane. (As is  known, this particular state  equation for the cosmic fluid leads to repulsive gravitation in conventional four-dimensional cosmology.) However, as a striking and perhaps unexpected result, we find that the presence of a positive energy density fluid on one brane leads to a {\it negative} energy density fluid on the other brane. This result is physically non-acceptable, and it makes one wonder about how far it is possible to give a physically meaningful interpretation of the two-brane scenario in general. One would expect that a two-brane, zero-tension, fluid-containing system should be a simple and physically meaningful system, but the formalism shows that it actually is not.

(2)  We wish to analyze the localization of gravity on Friedmann-Robertson-Walker type branes embedded in either $AdS_5$ or $dS_5$ bulk space, discussing in particular the lower limit of the fluid energy density on the first (TeV) brane when the effective four-dimensional cosmological constant is positive. We also solve the governing equation for the perturbed metric, and show in the limiting case of small Kaluza- Klein  mass $m$ that they do not modify Newton's inverse square law. This analysis extends the analysis for one single brane, in the form as given in Ref.~\cite{brevik02}. (For an earlier indirect proof of such a localization, see Ref.~\cite{nojiri01a}.)

There are of course many facets of brane dynamics that are not covered here. Thus we assume, as mentioned,  that  the component $T_{ty}$ of the five-dimensional energy-momentum tensor is zero. Bulk gravitons produced by the brane matter fluctuations have recently been analyzed in Refs.~\cite{hebecker01,langlois02a,kiritsis03}. Another simplification worth mentioning is that we assume both fluids to be ideal, {\it i.e.,} non-viscous. The theory of viscous fluids in a brane context has recently been investigated in Refs.~\cite{brevik03,chen01,harko03}.

In the next section we establish the Einstein equations for the case of one brane and derive, by means of the gauge conditions, the first Friedmann equation showing the presence of the $\rho^2$ term which is so characteristic for five-dimensional cosmology. In Sec. III we consider two parallel branes at a fixed separation $R$, each brane containing an isotropic fluid. As mentioned, $\sigma =0$ is assumed.  Setting the "dark radiation term" in the Friedmann equation equal to zero, we give the formal solutions for the components of the metric tensor, in the $dS_5$ case as well as in the $AdS_5$ case. Explicit solutions are worked out in full when it is in addition assumed that $p_0=-\rho_0$ on the first brane. This equation is tantamount to assuming the brane to possess a cosmological constant, but to be otherwise fluid-free. It is shown how the Friedmann equation, together with the condition $T_{ty}=0$, make the two branes closely linked with each other: If $p_0=-\rho_0$ on the first brane, then necessarily $p_R=-\rho_R$ on the second brane also; however, with the notable property that $\rho_R <0$, as mentioned.  In Sec.~IV we consider the graviton localization problem, taking the horizon distance to be larger than the brane separation so that there is no point between the branes at which the metric component $g_{00}$ vanishes. Performing the analysis in full for the case close to RS fine-tuning, we show how the governing equations permit no solution for the perturbed metric in the bulk region. The gravitons are thus in this case bound to the branes.

\section*{\label{II}II.  Einstein's Equations. One Single Brane}

It will be helpful first to establish the basic formalism, in the presence of one single brane lying at $y=0$. The metric will be taken in the form
\begin{equation}
ds^2= -n^2(t,y)dt^2+a^2(t,y)\gamma_{ij}dx^idx^j +dy^2,
\label{1}
\end{equation}
where
\begin{equation}
\gamma_{ij}(x)\equiv \xi^{-2}\delta_{ij}=\left(1+\frac{1}{4}k\,\delta_{mn}\,x^mx^n\right)^{-2}\delta_{ij},
\label{2}
\end{equation}
and $k=-1,0,1$. The quantities $n(t,y)$ and $a(t,y)$ are determined from Einstein's equations. (Note that the coordinate $y$ is the same as Randall-Sundrum's  $r_c\,\phi$, where $r_c$ is a measure of the fifth dimension and $\phi$ is a nondimensional coordinate lying in the interval $0 \leq \phi  \leq \pi$.)

The five-dimensional Einstein equations are
\begin{equation}
R_{M N}-\frac{1}{2}g_{MN}R +g_{MN}\Lambda=\kappa^2T_{MN},
\label{2b}
\end{equation}
where the coordinate indices are numbered as $x^M=(t,x^1,x^2,x^3,y)$, and $\kappa^2=8\pi G_5$.

The components of the Ricci tensor in an orthonormal basis, designated with carets, are
\begin{equation}
R_{\hat{t}\hat{t}}=3\left( \frac{a'n'}{an}+\frac{\dot{a}\dot{n}}{an^3}-\frac{\ddot{a}}{an^2} \right)+\frac{n''}{n},
\label{3}
\end{equation}
\begin{equation}
R_{\hat{i}\hat{i}}=\frac{\ddot{a}}{an^2}-\frac{\dot{a}\dot{n}}{an^3}-\frac{a'n'}{an}+2\left[\frac{k}{a^2}+ \left(\frac{\dot{a}}{an}\right)^2
-\left(\frac{a'}{a}\right)^2\right]-\frac{a''}{a}\quad \rm{(no\; sum)},
\label{4}
\end{equation}
\begin{equation}
R_{\hat{y}\hat{y}}=-\frac{n''}{n}-\frac{3a''}{a},
\label{5}
\end{equation}
\begin{equation}
R_{\hat{t}\hat{y}}=3\left( \frac{\dot{a}n'}{an^2}-\frac{\dot{a}'}{an}\right).
\label{6}
\end{equation}
When expressed in a coordinate basis, Einstein's equations become
\begin{equation}
3\left\{ \left( \frac{\dot{a}}{a}\right)^2
-n^2 \left[\frac{a''}{a}
+\left(\frac{a'}{a} \right)^2 \right] +k\left(\frac{n}{a}\right)^2\right\}-\Lambda n^2=\kappa^2 T_{tt},
\label{8}
\end{equation}
\begin{align}
a^2\gamma_{ij} \Bigg\{& \frac{a'}{a}\left( \frac{a'}{a}+\frac{2n'}{n}\right) +\frac{2a''}{a}+\frac{n''}{n}\nonumber\\
&+\frac{1}{n^2}\Big[ \frac{\dot{a}}{a}\left(-\frac{\dot{a}}{a}+\frac{2\dot{n}}{n}\right)-\frac{2\ddot{a}}{a}\Big] -\frac{k}{a^2}+
\Lambda \Bigg\}
=\kappa^2 T_{ij},
\label{9}
\end{align}
\begin{equation}
3\left( \frac{\dot{a}}{a}\frac{n'}{n}-\frac{\dot{a}'}{a} \right)=\kappa^2T_{ty},
\label{10}
\end{equation}
\begin{equation}
3\Bigg\{\frac{a'}{a}\left(\frac{a'}{a}+\frac{n'}{n}\right)-\frac{1}{n^2}\Big[\frac{\dot{a}}{a}\left( \frac{\dot{a}}{a}-\frac{\dot{n}}{n}\right)
+\frac{\ddot{a}}{a}\Big]-\frac{k}{a^2}\Bigg\}+\Lambda=\kappa^2 T_{yy}.
\label{11}
\end{equation}
Overdots mean derivatives with respect to $t$, whereas primes mean derivatives with respect to $y$.

A remark on dimensions: If $k=\pm 1$, the spatial coordinate $x^i$ has to be nondimensional, implying that $a(t,y)$ has to carry the spatial dimension cm. Moreover, $t$ and $y$ are dimensional quantities. We may summarize the dimensions: $[x^i]=1,\,[a(t,y)]=[t]=[y]$= cm. This means that $[n(t,y)]=1$. It becomes natural to use the same conventions if $k=0$ also.

As energy-momentum tensor we take the form
\begin{equation}
T_{MN}=\delta(y)(-\sigma g_{\mu\nu}+\rho U_{\mu}U_\nu +p h_{\mu\nu})\delta_M^N\,\delta_N^\nu.
\label{12}
\end{equation}
This expression is composed of two parts: one part which in an orthonormal frame means $T_{\hat t\hat t}=\delta(y)\,\sigma, \; T_{\hat i \hat j}=-\delta(y)\,\sigma$, implying the usual equation of state $p=-\sigma$ for a cosmic brane \cite{vilenkin81}, and another part which describes the energy-momentum for an ideal fluid. We have introduced here the projection operator $h_{\mu\nu}=g_{\mu\nu}+U_\mu U_\nu$. The bulk space itself ($y\neq 0$) does not contribute to $T_{MN}$.

We work henceforth in an orthonormal frame, in which $U^\mu=(1,0,0,0)$. With the notation $a_0(t)=a(t,y=0)$ and similarly for $n_0(t)$, we have on the brane
\begin{equation}
ds^2=-dt^2+a_0^2(t)\,\gamma_{ij}(x)dx^idx^j,
\label{13}
\end{equation}
where we have imposed the gauge condition $n_0(t)=1$, which means that the proper time on the brane is taken as the time coordinate.

The gauge conditions at $y=0$ are handled as in earlier papers - cf., for instance, Refs.~\cite{binetruy00a,brevik02,langlois03} - and lead to the equation
\begin{equation}
\frac{[a']}{a_0}=-\frac{1}{3}\kappa^2(\sigma+\rho),
\label{14}
\end{equation}
where $[a']=a'(0^+)-a'(0^-)$ is the jump across $y=0$, and similarly to
\begin{equation}
[n']=\frac{1}{3}\kappa^2 (-\sigma +2\rho+3p).
\label{15}
\end{equation}

We will moreover assume that there is no energy flux in the $y$ direction:
\begin{equation}
T_{ty}=0.
\label{16}
\end{equation}
We derive after some calculation the first Friedmann equation
\begin{equation}
H_0^2=\lambda-\frac{k}{a_0^2}+\frac{1}{18}\kappa^4\sigma \rho+\frac{1}{36}\kappa^4\rho^2+\frac{C}{a_0^4},
\label{17}
\end{equation}
where the quantity
\begin{equation}
\lambda=\frac{1}{6}\Lambda+\frac{1}{36}\kappa^4\sigma^2
\label{18}
\end{equation}
is interpreted as an effective four-dimensional cosmological constant in the five-dimensional theory. Subscript zero refers to the brane position; $C$ is an integration constant; $H_0=\dot{a}_0/a_0$ is the Hubble parameter.

It should be mentioned that the expression (\ref{17}) can be obtained from the corresponding expression pertaining to a brane containing no fluid at all, if we make the substitution $\sigma \rightarrow \sigma +\rho$. This substitution naturally follows from an inspection of the $(t,t)$ and $(t,y)$ components of Einstein's equations and the junction conditions.

As an example, let us reproduce from \cite{brevik02} the solution in the AdS case, $\Lambda <0$:
\[a^2(t,y)=\frac{1}{2}a_0^2 \left( 1+\frac{\kappa^4 \sigma^2}{6\Lambda}\right)+\frac{3C}{\Lambda a_0^2}\]
\begin{equation}
+\left[\frac{1}{2}a_0^2 \left(1-\frac{\kappa^4 \sigma^2}{6\Lambda}\right)-\frac{3C}{\Lambda a_0^2}\right]
\cosh (2\mu\, y)-\frac{\kappa^2 \sigma}{6\mu} \,a_0^2 \sinh (2\mu |y|),
\label{19}
\end{equation}
where $\mu=\sqrt{-\Lambda/6}$. The subsequent expression for $a_0(t)$ was however given incorrectly in \cite{brevik02}, so let us correct it here. It should read
\begin{equation}
a_0(t)=\frac{1}{2\sqrt{\lambda}\,f(t)}\left[f^4(t)-4\lambda\,C+2kf^2(t)+k^2\right]^{1/2},
\label{20}
\end{equation}
where
\begin{equation}
f(t)=e^{\sqrt{\lambda}(t+c_0)},
\label{21}
\end{equation}
$c_0$ being a new integration constant.

Before considering the two-brane geometry, let us briefly comment on the Friedmann equation, Eq.~(\ref{17}). First, we see that the condition $\lambda=0$, or $\sigma=6\mu/\kappa^2$, is for $k=0,\,\Lambda <0$ the same as the Randall-Sundrum fine-tuning condition. Moreover, the third term on the right hand side of Eq.~(\ref{17}), being linear in $\rho$, is of the same kind as in four-dimensional cosmology. The quadratic fourth term has however no counterpart in 4D theory, and becomes influential only in the case of very high energy. To get an idea about the magnitude of the high energy correction, let us assume the simple case where $\lambda=k=C=0$, so that the Friedmann equation reduces to $H_0^2=\kappa^4 \rho^2/36$. Together with the energy conservation equation
\begin{equation}
\dot{\rho}+3H_0(\rho+p)=0
\label{22}
\end{equation}
and the equation of state
\begin{equation}
p=(\gamma -1) \rho,
\label{23}
\end{equation}
 with $\gamma$ a constant, we then get
\begin{equation}
a_0(t) \propto t^{\frac{1}{3\gamma}},
\label{24}
\end{equation}
instead of the conventional 4D expression
\begin{equation}
a_0(t) \propto t^{\frac{2}{3\gamma}}.
\label{25}
\end{equation}
This means that the expansion of the universe is slowed down in the 5D case.

The last term $C/a_0^4$ in Eq.~(\ref{17}) behaves like a radiation term - cf., for instance, the discussion in Refs.~\cite{langlois03,langlois03a} - and is called the dark radiation term. From the viewpoint of
 the AdS/CFT correspondence, the dark radiation can be regarded as CFT radiation \cite{brevik02,gubser01,padilla02}. (The theory in \cite{padilla02} was generalized in \cite{gregory02}.) This term can be omitted in the various epochs of the history of the universe, except in the radiation epoch.

\section*{\label{III}III.  Two Flat Branes}

\subsection*{A. The Friedmann Equations}

Consider now the two-brane configuration, in which the fifth dimension $y$ is compactified on an orbifold $S^1/Z_2$ of radius $R/\pi$, with $-R \leq y \leq R$. The orbifold fixed points at $y=0$ and $y=R$ are the locations of the two three-branes, which form the boundary of the 5D spacetime. If $\Lambda <0$, the spacetime between the two branes located at $y=0$ and $y=R$ is a slice of $AdS_5$ geometry. As usual, we identify the first brane at $y=0$ with the high energy Planck brane, whereas the second brane at $y=R$ is the low energy TeV brane.

The energy-momentum tensor describes matter on the branes:
\begin{equation}
T_M^N=\delta(y)\,{\rm diag }(-\rho_0,p_0,p_0,p_0,0)+\delta(y-R)\,{\rm diag} (-\rho_R,p_R,p_R,p_R,0).
\label{26}
\end{equation}
We make henceforth the assumption that the brane tension on either brane is zero:
\begin{equation}
\sigma=0.
\label{27}
\end{equation}
This assumption restricts the scope of our theory. Specifically: In the usual setting when there is a brane tension, but  no fluid, the situation is still  encompassed by our theory since this corresponds simply to choosing $p=-\rho$ as state equation.  Such a  `vacuum' fluid is physically equivalent to a cosmological constant. However, the general case is when $\sigma \neq 0$ and when, in addition, there are brane fluids endowed with a general value of $\gamma$ in the state equation. Such a general situation is outside the scope of the present paper.

We adopt the same metric as before, implying that the Einstein tensor does not change. Integrating the $(t,t)$ or $(y,y)$ components of Einstein's tensor by making use of the $(t,y)$ component we obtain
\begin{equation}
\left(\frac{\dot{a}}{na}\right)^2=\frac{1}{6}\Lambda-\frac{k}{a^2}+\left(\frac{a'}{a}\right)^2+\frac{C}{a^4}
\label{28}
\end{equation}
in the bulk. As junction conditions we now have, from Eq.~(\ref{8}),
\begin{equation}
\frac{[a']_0}{a_0}=-\frac{1}{3}\kappa^2\,\rho_0,\quad \frac{[a']_R}{a_R}=-\frac{1}{3}\kappa^2\,\rho_R,
\label{29}
\end{equation}
and similarly from Eq.~(\ref{9})
\begin{equation}
\frac{[n']_0}{n_0}=\frac{1}{3}\kappa^2\,(2\rho_0+3p_0),\quad \frac{[n']_R}{n_R}=\frac{1}{3}\kappa^2
(2\rho_R+3p_R).
\label{30}
\end{equation}
>From the $Z_2$ symmetry and the continuity of $a$ we have $[a']_0=a'(0^+)-a'(0^-)=2a'(0^+)\equiv 2a'(0)$, and similarly $[a']_R=a'(R^+)-a'(R^-)=-2a'(R^-)\equiv -2a'(R).$ When this is used in Eq.~(\ref{29}) we obtain for the Friedmann equation on each brane, Eq.~(\ref{28}), by choosing $n_0(t)=1$ on the first (Planck) brane,
\begin{equation}
H_0^2=\frac{1}{6}\Lambda-\frac{k}{a_0^2}+\frac{1}{36}\kappa^4\,\rho_0^2+\frac{C}{a_0^4},
\label{31}
\end{equation}
\begin{equation}
{\cal H}_R^2=\frac{1}{6}\Lambda-\frac{k}{a_R^2}+\frac{1}{36}\kappa^4\,\rho_R^2+\frac{C}{a_R^4},
\label{32}
\end{equation}
with
\begin{equation}
{\cal H}_R=\frac{da_R/d\tau}{a_R}=\frac{1}{n_R}\frac{\dot{a}_R}{a_R}
\label{33}
\end{equation}
being the Hubble parameter on the second (TeV) brane. Note that the cosmological time on the first brane is still denoted by $t$, whereas the cosmological time element on the second brane is $d\tau= n_R\, dt$.

The condition $T_{ty}=0$ on the first brane yelds, when account is taken of Eq.~(\ref{10}),
\begin{equation}
\dot{\rho}_0+3H_0\,(\rho_0+p_0)=0.
\label{34}
\end{equation}
Formally, this is in agreement with the one-brane result, Eq.~(\ref{22}). Similarly, the same condition applied on the second brane yields
\begin{equation}
\frac{d\rho_R}{d\tau}+3{\cal H}_R(\rho_R+p_R)=0.
\label{35}
\end{equation}

\subsection*{B.   Solving for the Metric}

In view of the choice $n_0(t)=1$, the condition $T_{ty}=0$ implies that the relation
\begin{equation}
n(t,y)=\frac{\dot{a}(t,y)}{\dot{a}_0(t)}
\label{36}
\end{equation}
follows from the $(t,y)$ component of Einstein's  equations. From the $(t,t)$ component of the same equations it follows that
\begin{equation}
(\dot{a}_0)^2-(aa')'+k=\frac{1}{3}\Lambda\,a^2
\label{37}
\end{equation}
in the bulk.

Let us assume that the constant $C$ in Eqs.~(\ref{31}) and (\ref{32}) is zero. The calculation of $a(t,y)$ becomes analogous to that of the one-brane case. We consider first a $dS_5$ bulk, $\Lambda >0$.  With the definition $\mu_d=\sqrt{\Lambda/6}$ we obtain, when taking the $Z_2$ symmetry into account,
\begin{equation}
a(t,y)=a_0(t)\left( \cos(\mu_d\,y)-\frac{\kappa^2\,\rho_0}{6\mu_d}\sin(\mu_d\,|y|)\right).
\label{38}
\end{equation}
We have here taken the positive square root of $a^2(t,y)$.
Note that at this stage we cannot factorize $a$ as $a(t,y)=a_0(t)A(y)$. The reason is the possible time dependence of the energy density $\rho_0$.

Using Eq.~(\ref{36}) we can also determine $n(t,y)$:
\begin{equation}
n(t,y)=\cos(\mu_d\,y)-\frac{\kappa^2}{6\mu_d}\left(\rho_0+\frac{\dot{\rho}_0}{H_0}\right)\sin(\mu_d\,|y|).
\label{39}
\end{equation}
For an $AdS_5$ bulk ($\Lambda<0$) we similarly obtain
\begin{equation}
a(t,y)=a_0(t)\left(\cosh(\mu y)-\frac{\kappa^2\,\rho_0}{6\mu}\sinh(\mu\,|y|)\right),
\label{40}
\end{equation}
\begin{equation}
n(t,y)=\cosh(\mu y)-\frac{\kappa^2}{6\mu}\left(\rho_0+\frac{\dot{\rho}_0}{H_0}\right)\sinh(\mu\,|y|),
\label{41}
\end{equation}
with $\mu=\sqrt{-\Lambda/6}$.

The so far undetermined quantity $a_0(t)$ determined from Eq.~(\ref{31}) depends on $\rho_0, \Lambda$, and $k$. We will not solve for $a_0(t)$ in general, but specialize henceforth to the case when
\begin{equation}
p_0=-\rho_0= {\rm constant}
\label{42}
\end{equation}
on the first brane.
As mentioned earlier, this is tantamount to assuming the first brane to possess a cosmological constant, but being otherwise fluid-free. It is now natural to define the effective 4D cosmological constant as
\begin{equation}
\lambda_0=\frac{1}{6}\Lambda+\frac{1}{36}\kappa^4 \rho_0^2;
\label{43}
\end{equation}
this replaces the previous definition in Eq.~(\ref{18}). We see that the Friedmann equation (\ref{31}) is identical with the one-brane equation (\ref{17})
(with $\sigma=0$). This allows us to make use of the solutions obtained for one brane \cite{brevik02}, replacing $\lambda$ with $\lambda_0$. For $\lambda_0>0$,
\begin{eqnarray}
a_0(t)=\left\{ \begin{array}{ll}
e^{\sqrt{\lambda_0} t}                                                  & ,k=0 \\
\frac{1}{\sqrt{\lambda_0}}\cosh(\sqrt{\lambda_0}\,t+\alpha_1)     & ,k=1 \\
\frac{1}{\sqrt{\lambda_0}}\sinh(\sqrt{\lambda_0}\,t+\alpha_2)    & ,k=-1,
\end{array}
\right.
\label{44}
\end{eqnarray}
where $\alpha_1$ and $\alpha_2$ are integration constants.

The case of $\lambda_0<0$ is possible only for $AdS_5$ and $k=-1$. We then get
\begin{equation}
a_0(t)=\frac{1}{\sqrt{-\lambda_0}}\sin(\sqrt{-\lambda_0}\,t),\quad k=-1.
\label{45}
\end{equation}
The assumed constancy of $\rho_0$ now makes it possible to factorize the metric:
\begin{equation}
a(t,y)=a_0(t)A(y),\quad n(t,y)=A(y);
\label{46}
\end{equation}
cf. Eqs.~(\ref{38}),(\ref{39}) or (\ref{40}),(\ref{41}). This product form separates Eq.~(\ref{28}) into
\begin{equation}
\left( \frac{\dot{a}_0}{a}\right)^2+\frac{k}{a_0^2}=(A')^2+\frac{1}{6}\Lambda\, A^2 =\lambda_0,
\label{47}
\end{equation}
where the last equality follows from Eq.~(\ref{31}). Evaluation of Eq.~(\ref{47}) on the second brane yields
\begin{equation}
\Big[ \frac{1}{6}\Lambda+\frac{1}{36}\kappa^4\,(2\rho_R+3p_R)^2 \Big]A^2(R)=\lambda_0.
\label{48}
\end{equation}
Thus, $(2\rho_R+3p_R)$ must be a constant. From this equation, even the weak equation of state $p_R=w \rho_R $ for the fluid, with $w$ a constant, would suffice to ensure that $\rho_R$ = constant. However, there is an additional condition on the system, namely $T_{ty}=0$, which makes the restriction on the equation of state stronger: From Eq.~(\ref{35}) it follows that we must have
\begin{equation}
p_R=-\rho_R.
\label{49}
\end{equation}
This is as we would expect, from analogy with Eq.~(\ref{42}). The two branes are linked, via the gap in the fifth dimension.  Moreover, it is seen from Eq.~(\ref{48}) that $\lambda_R$ (defined analogously to $\lambda_0$ in Eq.~(\ref{43})) and $\lambda_0$ are of the same sign, and related through $A^2(R)$:
\begin{equation}
\lambda_R\,A^2(R)=\lambda_0.
\label{50}
\end{equation}
The branes are thus both $AdS_4$ ($\lambda<0$), $M_4$ ($\lambda=0$), or $dS_4$ ($\lambda>0$). The gap width $R$ is a function of the bulk cosmological constant $\Lambda $ and the brane energy densities $\rho_0$ and $\rho_R$. The quantity $A(R)$ is found from $A(y)$, using Eq.~(\ref{40}) and  assuming $\Lambda<0$:
\begin{equation}
A(y)=\frac{\sqrt{\lambda_0}}{\mu}\sinh[\mu(y_H-|y|)].
\label{51}
\end{equation}
Here $y_H \, (>0)$ is the horizon, defined by
\begin{equation}
\tanh(\mu \,y_H)=\frac{6\mu}{\kappa^2 \rho_0}, \quad {\rm or} \quad \sinh(\mu\, y_H)=\frac{\mu}{\sqrt{\lambda_0}}.
\label{52}
\end{equation}
Thus $A(0)=1$ as it should.  From  Eq.~(\ref{51}) it is seen that $A(y)$ decreases monotonically with increasing distance $|y|$ from the first brane.  Equation (\ref{50}) thus implies that $\lambda_R> \lambda_0$, which in turn implies that
\begin{equation}
|\rho_R| > |\rho_0|.
\label{52a}
\end{equation}
We can at this stage conclude:  The existence of a finite gap between the branes implies that the {\it magnitude} of the energy density on the second brane is higher than the magnitude of the energy density on the first brane.

However, is this picture realistic physically? We have so far assumed, in conformity with usual practice, that the first brane possesses a positive tensile stress (negative pressure). Thus $\rho_0 =-p_0 >0$. Let us go  back to the junction conditions (\ref{30}) and write them, in view of the separability condition (\ref{46}), as
\begin{equation}
\frac{A'(0^+)}{A(0)}=-\frac{1}{6}\kappa^2 \rho_0, \quad \frac{A'(R^-)}{A(R)}=\frac{1}{6}\kappa^2 \rho_R.
\label{52b}
\end{equation}
Thus, the property $A'(R^-) <0$ implies that the fluid energy density on the second brane becomes {\it negative}:
\begin{equation}
\rho_0 >0  \Rightarrow \rho_R < -\rho_0.
\label{52c}
\end{equation}
This is physically non-acceptable; the whole picture about a fluid residing on the second brane breaks down. We cannot accept that there is a negative energy density for a fluid in its rest inertial frame.

One may ask: Our considerations above apparently gave the first brane a privileged status. From a democratic point of view, should not the same behavior be found if we instead started out with a situation where the energy density $\rho_R$ on the second brane were positive? The answer actually turns out to be affirmative. Specifically, instead of the expression (51) for $A(y)$ we may alternatively write
\begin{equation}
A(y)=\frac{\sqrt{\lambda_0}}{\mu}\sinh [\mu (y_H+|y|)],
\label{52d}
\end{equation}
where $y_H$ is the same positive quantity as before, and where now
\begin{equation}
\tanh (\mu y_H)=-\frac{6\mu}{\kappa^2\rho_0}.
\label{52e}
\end{equation}
The junction conditions (\ref{52b}) are the same as before. Since now $A(0)=1,\, A(R)>1$,  it follows that $|\rho_0|>|\rho_R|$. Accordingly,
\begin{equation}
\rho_R >0 \Rightarrow \rho_0<-\rho_R.
\label{52f}
\end{equation}
Whereas the above argument was carried out for $\Lambda <0$, a similar analysis for the case $\Lambda >0$ leads to the same conclusion. Thus, the property of Eqs.~(\ref{52c}) and (\ref{52f}) hold for either sign of $\Lambda$, and is a characteristic property of the $dS_4$ class ($\lambda >0$).

Note that this difficulty with interpretation still persists in the simple case of fine tuning. Let us put $\lambda_0=0$, and also take $k=0$. Then,
\begin{equation}
A(y)=e^{-\mu |y|},
\label{52g}
\end{equation}
which implies that also $\lambda_R=0$ in view of Eq.~(\ref{50}). We get
\begin{equation}
\rho_0=\frac{\sqrt{-6\Lambda}}{\kappa^2}, \quad \rho_R=-\frac{\sqrt{-6\Lambda}}{\kappa^2}.
\label{52h}
\end{equation}
The expression for $\rho_0$ is acceptable (we still assume $\Lambda <0$), but that for $\rho_R$ is not.

Physically speaking, the above perhaps surprising properties  seem to reflect the peculiar behavior encountered in the standard RS setting where the branes are taken to possess tensions. Conventionally, when assuming no fluids to be present on the branes, one finds that a {\it positive} tensile stress $\sigma$ on the first brane is accompanied by a {\it negative} tensile stress on the second brane.  In the present theory, we have put $\sigma =0$ (cf. Eq.~(\ref{27})). Instead of imagining  the second brane as a negative tensile stress brane, we find it to be endowed with a negative energy density fluid. The fluid brane picture is thus physically  more restrictive than the tensile stress picture. At present it is hardly possible to decide which of the descriptions is the most realistic one.

\subsection*{C. Stability of Configuration}

Allowing the interbrane distance to depend on time, {\it i.e.} $R = R(t)$, it is of interest to study the stability of the two-brane system. In the case of an $AdS_5$ bulk, this has been performed in \cite{langlois02}. It is found that a configuration of $dS_4$ branes is unstable whereas in the case of $AdS_4$ and $M_4$ ($\lambda \leq 0$) the interbrane distance remains finite. The instability of the interesting case of positive $\lambda$'s is a point of concern, but may also be viewed as a feature of cyclic universe models, such as in \cite{steinhardt01}. However, the configuration could also be stabilized by introducing a bulk scalar field, as done by Goldberger and Wise \cite{goldberger01} for the fine-tuned case of $\lambda_0 = \lambda_R = 0$. This would require including a Klein-Gordon field described by the following action:
\begin{align}
 \label{gwaction}
 S_b & = \frac{1}{2}\int d^4x \int_{-R}^{R} dy \sqrt{-g} (g^{MN}\partial_M \Phi \partial_N \Phi - m^2 \Phi^2), \\
 S_0 & = -\int d^4x \int_{-R}^{R} dy \sqrt{-g} \ l_0 (\Phi^2 -v_0^2)^2 \delta(y), \\
 S_R & = -\int d^4x \int_{-R}^{R} dy \sqrt{-g} \ l_R (\Phi^2 -v_R^2)^2 \delta(y-R) ,
\end{align}
where $S_b$ is the bulk term and $S_{0,R}$ are interaction terms on the branes, and $l_{0,R}$ and $v_{0,R}$ are constants. One may neglect the impact on the background metric by this addition, as argued in \cite{goldberger01}. By using the metric given by \eqref{51}, we arrive at the Euler-Lagrange equation for $\Phi$:
\begin{equation}
 \frac{d^2 \Phi}{dy^2} - \mathrm{sgn}(y) 4 \coth[\mu(y_H-|y|)] \frac{d \Phi}{dy} - m^2 \Phi = 0.
 \label{ELgw}
\end{equation}
Solving this differential equation for $\Phi$, one may insert the solution into the action and perform the $y$-integration. What is left may be considered as an effective potential for the distance between the branes. Performing this leads to complicated expressions, although using the same approximations as in \cite{goldberger01} results in some simplifications. With a suitable choice of parameters, one can show that the obtained effective potential has a minimum for finite $R$. This indicates that the interbrane distance in the $dS_4$-case can be stabilized by introducing a bulk field.

\section*{\label{IV} IV.  Localization of Gravity}

The RS scenario provides corrections to Newton's inverse square law. This law is experimentally verified for distances larger than about 200 $\mu$m \cite{hoyle01}. To reproduce the inverse square law to a satisfactory accuracy, gravity (or the graviton) has to be localized on our three-brane. The cases where $\lambda=0$ and $\lambda<0$ have been studied earlier. In Ref.~\cite{brevik02}, localization was shown to be possible in the case of a single $dS_4$ brane $(\lambda>0)$ embedded in either a $dS_5$ or an $AdS_5$ bulk. Let us investigate here which changes in formalism are caused by the presence of two flat branes. As above, we assume the equation of state   $p_0=-\rho_0,\; (\rho_0={\rm const}>0),$  on the first brane. Similarly, we assume $p_R=-\rho_R={\rm const}$ on the second brane. The negativity problem  for $\rho_R$ here does not come into play.

We assume $\lambda_0>0$ and $\lambda_R>0$, {\it i.e.}, $dS_4$ branes, and let the metric be perturbed as follows:
\begin{equation}
ds^2=-n^2(t,y)dt^2+a^2(t,y)\left[ \gamma_{ij}(x^i)+h_{ij}(x^\mu)\right]\,dx^idx^j + dy^2.
\label{53}
\end{equation}
As usual, we identify the transverse and traceless component $h$ of the perturbation $h_{ij}$with the graviton on the brane. Thus $h_i^i=0,\,\nabla_jh^{ij}=0$. For $k=0$, these conditions lead to the linearized equation for $h$,
\begin{equation}
\nabla_M\nabla^M h=0.
\label{54}
\end{equation}
Our assumption $C=0$ implies that the metric coefficients are separable as in Eq.~(\ref{46}). A Kaluza - Klein expansion for $h$,
\begin{equation}
h=\int dm\,\phi_m(t,x^i)\,\Phi(m,y),
\label{55}
\end{equation}
permits Eq.~(\ref{54}) to separate into a four-dimensional part for $\phi_m$ and an $y$- part for $\Phi$. We give the equation for $\Phi$ here:
\begin{equation}
\Phi''+\frac{4A'}{A}\Phi'+\frac{m^2}{A^2}\Phi=0.
\label{56}
\end{equation}
This equation holds also if $k=\pm 1$.

The governing equation (\ref{56}) for the perturbed metric can be transformed into a Schr\"{o}dinger-like equation
\begin{equation}
-u''(z)+V(z)u(z)=m^2u(z)
\label{57}
\end{equation}
(prime meaning derivative with respect to the argument shown), where $u(z)=A^{3/2}(y)\,\Phi(y),\; dy/dz=A(y)$. The potential $V$ when expressed in terms of $y$ is
\begin{equation}
V(y)=\frac{9}{4}(A'(y))^2+\frac{3}{2}A(y)A''(y).
\label{58}
\end{equation}
We recall that $A(y)$ is different according to whether $\Lambda>0$ or $\Lambda<0$ (though $\lambda$ is assumed positive in both cases). Let us henceforth assume $\Lambda<0$, so that Eqs.~(\ref{51}) and (\ref{52}) hold. The potential in Eq.~(\ref{58}) becomes
\[
V(y)=\frac{9}{4}\lambda_0\cosh^2[\mu (y_H-|y|)]+\frac{3}{2}\lambda_0\sinh^2[\mu(y_H-|y|)] \]
\begin{equation}
-\frac{1}{2}\kappa^2 [\rho_0 \,\delta(y)+\rho_R \frac{\lambda_0}{\lambda_R}\,\delta_R(|y|-R)].
\label{59}
\end{equation}
With
\begin{equation}
z=\frac{{\rm sgn}(y)}{\sqrt{\lambda_0}}\ln \left[ \coth \frac{\mu(y_H-|y|)}{2}\right]
\label{60}
\end{equation}
we have
\begin{equation}
A(z)=\frac{\sqrt{\lambda_0}}{\mu\, \sinh(\sqrt{\lambda_0}\,|z|)},
\label{61}
\end{equation}
and we can express the potential $V(z)$ in terms of $z$:
\[ V(z)=\frac{15}{4}\lambda_0\left[ \frac{1}{\sinh^2(\sqrt{\lambda_0}\,z)}+\frac{3}{5}\right] \]
\begin{equation}
-\frac{1}{2}\kappa^2[\rho_0\,\delta(|z|-z_0)+\rho_R \sqrt{\frac{\lambda_0}{\lambda_R}}\,\delta(|z|-z_R)].
\label{62}
\end{equation}
Here we have taken into account that $\delta (y-R)=\sqrt{\lambda_R/\lambda_0}\,\delta (z-z_R)$.
The expression (\ref{62}) corresponds to Eq.~(57) in Ref.~\cite{brevik02}, with the addition of an extra delta function term at $z=z_R$. The potential in Eq.~(\ref{62}) is of the volcano type \cite{ghoroku01}, with delta functions at the two boundaries. The appearance of boundary conditions at $z=z_0$ and $z=z_R$ implies that the energy spectrum becomes discrete.

It is of interest to explore the physical aspects of the present formalism more closely. First, our assumption about a $dS_4$ brane $(\lambda>0)$ embedded in an $AdS_5$ bulk $(\Lambda<0)$ implies according to Eq.~(\ref{43})
\begin{equation}
\kappa^2\,\rho_0 >6\mu.
\label{63}
\end{equation}
The positions of the two branes when expressed in terms of the $z$ coordinate are
\begin{equation}
z_0=\frac{1}{\sqrt{\lambda_0}}
{\rm arcsinh} \frac{\sqrt{\lambda_0}}{\mu},
\label{64}
\end{equation}
\begin{equation}
z_R=\frac{1}{\sqrt{\lambda_0}}\ln \left[ \coth \frac{\mu (y_H-R)}{2}\right].
\label{65}
\end{equation}
Equation (\ref{52}) yields aøways a real value for the horizon distance $y_H$. This behaviour is the same as in the one-brane case. The added complexity in the two-brane case is that the horizon may, or may not, lie in between the branes. We shall assume, as seems most natural physically, that it is the second option which is realized in nature. We thus put henceforth  $R<y_H$. Then the metric components do not become zero anywhere between the branes. Integration of Eq.~(\ref{57}) across the branes yields the following boundary conditions:
\begin{equation}
\frac{du}{dz}\Bigg|_{z_0+}=-\frac{1}{4}\kappa^2\,\rho_0\,u(z_0),
\quad \frac{du}{dz}\Bigg|_{z_R-}=\frac{1}{4}\kappa^2\,\rho_R \sqrt{\frac{\lambda_0}{\lambda_R}}\,u(z_R).
\label{66}
\end{equation}
The solution of Eq.~(\ref{57}) in the bulk is
\begin{equation}
u(z)=c_1Y^{-id}{_2F_1}(a,b;c;-Y)+c_2Y^{id}{_2F_1}(a',b';c';-Y),
\label{67}
\end{equation}
where ${_2F_1}$ is Gauss' hypergeometric function, $c_1$ and $c_2$ are constants, and
\begin{equation}
d=\frac{1}{4}\sqrt{-9+\frac{4m^2}{\lambda_0}},\quad Y=\frac{1}{\sinh^2(\sqrt{\lambda_0}\,z)},
\label{68}
\end{equation}
\begin{equation}
a=-\frac{3}{4}-id,\quad b=\frac{5}{4}-id,\quad c=1-2id,
\label{69}
\end{equation}
\begin{equation}
a'=-\frac{3}{4}+id,\quad b'=\frac{5}{4}+id,\quad c'=1+2id.
\label{70}
\end{equation}
If $d$ is real ($m>\frac{3}{2}\sqrt{\lambda_0}$), the solution oscillates, whereas if $d$ is imaginary ($m<\frac{3}{2}\sqrt{\lambda_0}$), the two terms in the solution decrease/increase with $z$.

The general solution in Eq.~(\ref{67}) is rather complex, but there are special cases which are easy to analyze and which moreover are of physical interest. There are two values for the mass that are naturally singled out, namely $m=0$ and $m=\frac{3}{2}\sqrt{\lambda_0}$. Let us merely assume here that $m$ lies somewhere in this interval:
\begin{equation}
0 \leq m \leq \frac{3}{2}\sqrt{\lambda_0}\, ,
\label{71}
\end{equation}
and let us assume that $\lambda_0$ is positive but very small, so that we are close to the case of the RS fine-tuning. Also, we assume that the gap is narrow, when expressed in terms of the $z$ coordinate. Specifically, we assume
\begin{equation}
\frac{\lambda_0}{\sqrt{\mu}}\ll 1,\quad \sqrt{\lambda_0}\,z_R \ll 1.
\label{72}
\end{equation}
Then, Eq.~(\ref{57}) reduces to
\begin{equation}
u''(z)-\frac{15}{4\,z^2}u(z)=0,
\label{73}
\end{equation}
to leading order
in the bulk. The solution is
\begin{equation}
u(z)=c_1z^{5/2}+c_2z^{-3/2}.
\label{74}
\end{equation}
>From Eq.~(\ref{64}) we have for the position of the first brane
\begin{equation}
z_0=\frac{1}{\mu}.
\label{75}
\end{equation}
>From Eq.~(\ref{52}), $\tanh(\mu\, y_H)=1-\lambda_0/(2\mu^2)$ because $\mu\, y_H \gg 1$. Then, Eq.~(\ref{65}) yields
\begin{equation}
z_R=\frac{1}{\mu}\,e^{\mu R}.
\label{76}
\end{equation}
It turns out, however, that the solution (\ref{74}) does not satisfy the boundary conditions (\ref{66}) at $z=z_0$ and $z=z_R$ for any nonvanishing values of the constants $c_1$ and $c_2$. In the limiting case investigated, the perturbed metric does not propagate into the bulk. This is physically a natural result, since it does not modify Newton's inverse square law.

\section*{V. Concluding Remarks}

We have considered a static two-brane scenario, with a constant gap $R$ in the fifth dimension, and with an empty bulk except from the five-dimensional cosmological constant $\Lambda$. On the branes, situated at $y=0,R$, isotropic fluids with no viscosity were assumed present. No energy transport was assumed to take place from the branes into the bulk, {\it i.e.,} $T_{ty}=0$ for $y=0,R$. The brane tensions $\sigma$ were put equal to zero.

The Hubble parameters for the two branes are given by Eqs.~(\ref{31}) and (\ref{32}). They show the presence of a $\rho^2$ term, which is characteristic for five-dimensional cosmology. This term is negligible at low energies.  By contrast, the energy conservation equations (\ref{34}) and (\ref{35}) are of the same form as in conventional four-dimensional theory.

The "dark radiation term"  $C/a_0^2$ in Eqs.~(\ref{31}) and (\ref{32}) is related to the Weyl tensor. If $C=0$ (implying that also the Weyl tensor vanishes), then the metric components $a(t,y)$ and $n(t,y)$ are formally given by Eqs.~(\ref{38})-(\ref{41}) in the $dS_5$ case as well as in the $AdS_5$ case. Here any value of the  spatial curvature $k=-1,0,1$ is allowed, and the density $\rho_0$ may be time dependent. In the subsequent analysis in Sec.~III we required however $\rho_0$ to be time independent, and we assumed the equation of state for the fluid on the first brane to be $p_0=-\rho_0$, the latter assumption corresponding to the presence of a cosmological constant. In such a case, $a_0=a_0(t)$ is explicitly given by Eq.~(\ref{44}), and Eq.~(\ref{50}) shows how the two brane effective cosmological constants $\lambda_0$ and $\lambda_R$ are related. This implies, among other things,  that a fine-tuning ($\lambda_0=0$) of the first brane implies a fine-tuning ($\lambda_R=0$) of the second brane also.

A noteworthy result of the above analysis is that the "vacuum" equation of state $p_0=-\rho_0$ on the first brane implies that $p_R=-\rho_R$ on the second brane, but with $\rho_R<0$. (The inverse situation is analogous, showing the equivalence between the two branes.) This shows that a two-brane, zero-tension system with vacuum equations of state is  actually a problematic system physically.

It would be of interest to consider the more general situation in which the  condition about static branes is relaxed. There have recently been some investigations in this direction; cf., for instance, the paper of Maroto \cite{maroto03} analyzing the case where the branes are moving with constant velocity.

\bigskip

{\bf Acknowledgments}

\bigskip

We thank Sergei Odintsov and James Gregory for valuable information.

\end{document}